\documentclass[prl,twocolumn]{revtex4}

\usepackage{graphicx}
\usepackage{dcolumn}
\usepackage{bm}

\begin{document}


\title{Optical parametric oscillators in isotropic photonic crystals and cavities: \\
3D time domain analysis}

\author{C. Conti}
 \email{c.conti@ele.uniroma3.it}
 \homepage{http://optow.ele.uniroma3.it/opto_2002.shtml}
\author{A. Di Falco}%
\author{G. Assanto}
\affiliation{%
Nonlinear Optics and Optoelectronics Laboratory - NooEL \\
National Institute for the Physics of the Matter - INFM \\
Via della Vasca Navale, 84 - 00146 - Rome, Italy 
}%

\date{\today}

\begin{abstract}
We investigate optical parametric oscillations through four-wave mixing in resonant cavities and photonic crystals.
The theoretical analysis underlines the relevant features of the phenomenon and the role of the density of states.
Using fully vectorial 3D time-domain simulations, including both dispersion and nonlinear polarization, for the first time we address this process in a
face centered cubic lattice and in a photonic crystal slab. 
The results lead the way to the development of novel parametric sources in isotropic media.
\end{abstract}

\pacs{Valid PACS appear here}
\maketitle

Photonic crystals (PC) are synthetic structures designed for specific optical applications, realized by a 
(2 or 3D) periodic pattern embedded in a bulk material.
PC exhibit forbidden and allowed frequency bands of states for electromagnetic waves, depending on dielectric constants and geometric features, light polarization and direction of propagation. 
The analogy to solid state crystals in electronics is at the origin of their name \cite{Yablonovitch87,John87,JoannopoulosBook,SakodaBook}.
An alternative way to view PC is in terms of a periodic tight distribution of resonant cavities (RC),
such that the resulting system presents bands of states owing to mode splitting in the elemental RC.
A description of these structures exclusively in terms of free propagation within the allowed bands is rather limiting, especially in high-index contrast and finite-dimension PC.

The resonant response of PC is manifestly
relevant when considering nonlinear optical processes. By their very nature, in fact, the latter can
couple cavity modes, yielding a spectrum which is expected to significantly depend on the specific distribution of states.
If sufficient energy is available, 
the interplay between frequency (or mode) mixing and cavity effects may build up oscillations driven by the nonlinear gain. 
In PC, a high Q-factor \cite{Vuckovic99} entails 
low thresholds and, hence, a broad output spectrum which 
peaks at the most state-crowded frequencies.

While quadratic nonlinearities are the basis of most optical parametric oscillators (OPO), the process of four-wave mixing (FWM) in a cubic medium 
can trigger nonlinear gain and hence oscillations, as well. (see e.g. \cite{AgrawalBook,Brambilla91,Haelterman92,Coen01}, 
and references therein) Since it is present in all dielectrics, a cubic nonlinearity is an excellent
candidate for highly integrated parametric sources in PC, their realization crucially depending on materials.
Furthermore, the large density-of-states (DOS) characteristic of PC relaxes the selection rules that govern FWM parametric gain in resonant structures,
\cite{AgrawalBook} while providing several synchronized (phase-matched) interactions.   
The strict conditions that affect FWM in microresonator (without PC) have been pointed out in \cite{Spillane02}, where the 
observation of very low threshold (due to high Q-factor) Raman lasing has been reported. 

In this Letter we present a general model of OPO through FWM
in photonic crystals and cavities realized in isotropic materials (and
hence lacking a quadratic nonlinearity).
Such possibility is then verified by means of extensive vectorial 
time domain simulations, using a finite-difference-time-domain (FDTD) parallel algorithm which 
accounts for material dispersion in both the linear and nonlinear regimes.
To the best of our knowledge, these are the first fully comprehensive 3D vectorial time-domain  simulations of 
cavities and PC encompassing dispersion and a cubic optical response. 

Nonlinear optical phenomena in high index contrast 3D-PC were previously investigated numerically with reference to an instantaneous 
third-order nonlinearity, and to its role in distorting the band-structure \cite{Tran95,Lousse01}. 
A general theory was developed in \cite{Bath01} using the ``effective medium'' approach but
leaving aside the resonant features.
More recently, some experimental studies have been reported in PC \cite{Mazurenko03} (see also \cite{SlusherBook} for an updated 
review on nonlinear photonic crystals).

Hereby we focus, both theoretically and numerically, on frequency coupling associated to collective phenomena between several states in cavities.
Let us consider an RC, or a PC of finite extent, realized in an instantaneous isotropic medium. The PC is  
defined by a position-dependent refractive index $n ({\bf r})$,
in a region of volume $V$ and surface $S$ over which periodic boundary conditions are enforced.
Electric and magnetic fields can therefore be expandend in terms of cavity 
eigenfunctions, denoted by a generalized index $\mu$, according to 
\begin{equation}
\label{mainexpansion}
\begin{array}{l}
{\bf E}({\bf r},t)=\sum_\mu a_\mu(t) {\bf E}_\mu ({\bf r}) \exp(-i\omega_\mu t)\\
{\bf H}({\bf r},t)=\sum_\mu a_\mu(t) {\bf H}_\mu ({\bf r)} \exp(-i\omega_\mu t).
\end{array}
\end{equation}
In (\ref{mainexpansion}), as usual, the summation is extended over negative frequencies in order to make the fields
real-valued, with $-\mu$ corresponding to complex conjugation.
Applying the Lorentz reciprocity theorem to an instantaneously responding medium, one can derive the coupled mode 
equations \cite{Crosignani} for the amplitudes $a$:
\begin{equation}
\label{CBT}
\frac{da_\mu}{dt}=-\exp(i\omega_\mu t)\langle{\bf E}_\mu^* \cdot {\bf J}\rangle-\kappa_\mu a_\mu \text{,}
\end{equation} 
where the brackets denote the integral over $V$, and ${\bf J}$ is the current density associated to the nonlinear
polarization ${\bf P}_{NL}$ (i. e. ${\bf J}=\partial_t {\bf P}_{NL}$) or other coupling mechanisms, if present.
In (\ref{CBT}) we also included a loss coefficient $\kappa_\mu$, such that the quality factor is $Q_\mu=\omega_\mu/2\kappa_\mu$. 
\cite{Yariv66,Vuckovic99,MeystreBook}

For a third-order material we have:
$P_{NL}^l=\chi^{lmno}({\bf r})E^m E^n E^o $, the superscripts denoting the cartesian components, and having
omitted the sum over repeated symbols. 
Eqs. (\ref{CBT}) then become (the dot stands for time derivative):
\begin{equation}
\label{CBT2}
 \dot{a}_\mu+\kappa_\mu a_\mu=
-e^{i\omega_\mu t}\frac{d}{dt}
\sum_{\alpha\beta\gamma}g_{\mu,\alpha,\beta,\gamma}e^{-i(\omega_\alpha+\omega_\beta+\omega_\gamma)t}a_\alpha a_\beta a_\gamma\text{,}
\end{equation}
with coupling coefficients defined as the overlap integral between Bloch functions and the nonlinear response, i.e. 
$g_{\mu,\alpha,\beta,\gamma}=\langle \chi^{lmno}(E_\mu^l)^* E_\alpha^m E_\beta^n E_\gamma^o\rangle$.

Being interested in optical parametric oscillations due to a cubic nonlinearity, we consider a mode of amplitude
$a$ and frequency $\omega$ which, pumped by some external source, couples with two other modes of amplitues $a_+$ and $a_-$ at frequencies
$\omega_\pm$ respectively, such that $\omega_++\omega_-=2\omega$. 
Therefore, in the stationary regime (\ref{CBT2})
reads (for simplicity we neglect self- and cross-modulation terms):
\begin{equation}
\label{stationary}
\begin{array}{l}
\kappa a=2 i \omega g^* a^* a_+ a_-+p\\
\kappa_+ a_+=i \omega_+ g a^2 a_-^*\\
\kappa_- a_-=i \omega_- g a^2 a_+^*\text{,}
\end{array}
\end{equation}
with $g$ the relevant nonlinear coefficients and $p$ the pumping mechanism (see \cite{Yariv66}). The Manley-Rowe relations
are $P_+/\omega_+=P_-/\omega_-=-P/2\omega+pa^*$, with $P=2\kappa|a|^2$ the output power
corresponding to field amplitude $a$ (and similarly for $P_\pm$).  ``Below threshold''
 solutions of (\ref{stationary}) are $a_\pm=0$ and $a=p/\kappa$ (fixing $\arg(a)$), while
the threshold for oscillation is
\begin{equation}
\label{threshold}
|a|^2=\sqrt{\frac{\kappa_+\kappa_-}{\omega_+\omega_-}}\frac{1}{|g|}\text{,}
\end{equation}
or, in terms of the 
output photon flux at $\omega$: $P/\omega=1/2|g|Q\sqrt{Q_+ Q_-}$. 
As expected, large Q's reduce the threshold, while abobe threshold
the value of $a$ is clamped to
 (\ref{threshold}) and
\begin{equation}
\label{parametricenergy}
|a_\pm|^2=\frac{1}{4|g|Q}\sqrt{\frac{Q_\pm}{Q_\mp}}(\frac{p}{\kappa a}-1)\text{.}
\end{equation}
As in standard OPO's, the latter shows that the excess pumping 
($\Delta p=p/\kappa a-1$) is transferred to $\omega_\pm$.

Once the system is oscillating all the modes will vibrate at frequencies originated by four-wave mixing of $\omega$ and $\omega_\pm$. 
Henceforth, in the no-depletion approximation, $a$ and $a_\pm$ can be treated as source terms in (\ref{CBT2}).
If $\omega_\pm=\omega\pm\Delta\omega$, the generated frequencies $\omega_i$ are $3\omega$, $3\omega\pm\Delta\omega$,
$\omega\pm2\Delta\omega$, $2\omega\pm\Delta\omega$. 
Thereby, $a_\nu$ is the sum over all the FWM terms at each $\omega_i$. 
For instance, if $\omega_2\equiv\omega+\omega_+-\omega_-$, for any mode of order $\nu$ we have 
\begin{equation}
\label{modes}
\dot{a}_\nu+\kappa_\nu a_\nu=
i\omega_2 f_v(\omega,+\omega_+,-\omega_-)a a_-^* a_+ e^{i(\omega_\nu-\omega_2)t}\text{,}
\end{equation}
being $f_\nu(\omega,+\omega_+,-\omega_-)$ an effective nonlinear coefficient.
The stationary solution of (\ref{modes}) is 
\begin{equation}
\label{modenu}
a_\nu^{(\omega_2)}=\frac{i\omega_2 f_\nu(\omega,+\omega_+,-\omega_-)a a_-^* a_+}
{\kappa_\nu+i(\omega_\nu-\omega_2)}e^{i(\omega_\nu-\omega_2)t} \text{.}
\end{equation}

The energy at $\omega_2$ is $\mathcal{E}(\omega_2)=\sum_\nu |a_\nu^{(\omega_2)}|^2$,
which can be cast in the form
\begin{equation}
\label{energy}
 \mathcal{E}(\omega_2)=
\frac{\omega_2^2 |f_\nu(\omega_2)|^2_{\text{avg}}\Delta p^2}
{32|g|^3 Q^2 \sqrt{Q_+ Q_-}}
\int 
\frac{\rho(\omega)}{\kappa_{2}^2+(\omega-\omega_2)^2}d\omega \text{,}
\end{equation}
where we introduced an averaged nonlinearity $|f_\nu(\omega_2)|^2_{\text{avg}}$
and the DOS $\rho(\omega)$ of the cavity.\cite{SakodaBook} 
In deriving (\ref{energy}) we considered that, if $Q(\omega_2)\equiv Q_2$ 
 is sufficiently high, the relevant modes are those in the proximity of $\omega_2$.
If the DOS is smoothly varying (see \cite{Li01}), expression (\ref{energy})
states that the energy is approximately proportional to $\rho(\omega_2)$.

Increasing the pump fluence, the amplitudes $a_\pm$ become 
sources of other parametric processes, and more frequencies are generated.
At high powers the whole oscillation spectrum will resemble the DOS 
(with the obvious exception of those interactions for which $g=0$ 
due to symmetry). It will be peaked in regions where the states are denser, i.e. in the proximity of 
the photonic band-gap (PBG) for a PC or near cut-off for an RC.\cite{MeystreBook}

It is important to determine the bandwidth of the pump modes. The pump, at 
frequency $\omega_{in}$, transfers energy to each mode at $\omega=\omega_{in}-\delta$,
according to its Lorentzian lineshape.
We can account for this effect by writing the source term as $p e^{-i\delta t}$ and
solving the time-dependent coupled equations. This leads to replacing $\kappa$ and $\omega$ by
$\kappa+i\delta$ and $\omega_{in}$, respectively, in the first of (\ref{stationary}), and in corresponding ansatz's for $a_\pm$.
As a result, when $\delta\neq0$ the threshold value for $a$ 
increases, and the pump-bandwidth is determined by 
\begin{equation}
\label{bandwidth}
\frac{|p/\kappa|^2}{1+(\delta/\kappa)^2}=\sqrt{
\frac{\kappa_+ \kappa_-}{\hat{\omega}_+\hat{\omega}_-}}
\sqrt{1+\frac{\delta^2}{[(\kappa_++\kappa_-)/2]^2}}\text{,}
\end{equation}
with $\hat{\omega}_\pm=
\omega+2\delta\kappa_\pm/(\kappa_++\kappa_-)$. 
For small $\delta$, with $\kappa_\pm\cong\kappa$, from (\ref{bandwidth}) we have 
$|\delta|/\omega\cong\sqrt{\Delta p}/Q$. 
The total pump energy is the sum of the energies in each pump-mode:
the oscillator will be more efficient if the DOS is strongly peaked around $\omega_{in}$.

With the aim of demonstrating the feasibility of OPO's in isotropic PC/RC, we performed fully vectorial numerical simulations of nonlinear Maxwell
equations, with a code based on the FDTD approach for dispersive materials \cite{Young01}. 
Maxwell equations in vacuum were coupled with a nonlinear Lorentz oscillator (see Eqs. (\ref{Maxwell}) below), yielding the induced polarization ${\bf P}$ 
($P^2={\bf P}\cdot{\bf P}$)
in regions where material is present: 
\begin{equation}
\label{Maxwell}
\begin{array}{l}
\nabla\times{\bf E}=-\mu_0 \partial_t {\bf H}\\
\nabla\times{\bf H}=\epsilon_0 \partial_t {\bf E}+\partial_t {\bf P}\\
\partial_t^2{\bf P}+2\gamma_0\partial_t{\bf P}+\omega_0^2 f(P) {\bf P}=\epsilon_0(\epsilon_s-1)\omega_0^2 {\bf E}\text{.}
\end{array}
\end{equation}
In the algorithm, the Yee's grid \cite{Yee66} was used to enforce continuity between different media,
and uniaxial phase-matched layers (UPML) were adopted at the boundaries \cite{TafloveBook} (details will be provided
elsewhere).
Particular care is required in adopting a specific form of the Lorentz oscillator $f(P)$,  
with $f(P)=1$ describing a linear single-pole dispersive medium. 
For an isotropic Kerr material we followed the suggestion by Koga \cite{Koga99}, i.e. $f(P)=[1+(P/P_0)^2]^{-3/2}$ 
($P_0$ is a measure of the nonlinearity). 
For small $P/P_0$ the latter reproduces a Kerr response while also accounting for higher order terms. Compared to 
the standard Kerr $f(P)=1+\chi P^2$, the resulting algorithm is stable even near the Courant limit.\cite{TafloveBook} For the effective Kerr-law coefficient $n_2$ we chose $n_2=1.5\times10^{-17}m^2/W$, representative of an entire class of semiconductors \cite{Aitchison97}. This value gives $P_0\cong1$ (MKS units).
Because of the significant computational resources needed for such a numerical approach, we parallelized the algorithm.
\footnote{The code runs on the IBM-SP4 system at the Italian Interuniversity Consortium for Advanced Calculus (CINECA),  
and on the BEOWULF cluster at NooEL}

A proper test-bed for our theoretical analysis is provided by a dispersive medium with non-instantaneous 
linear and nonlinear responses, modeled by (\ref{Maxwell}) and solved numerically with no-approximations. 
We computed the response of a Face-Centered-Cubic (FCC) lattice (of period $\Lambda$) of air-spheres (radius $r=0.3535\Lambda$) embedded in a dielectric. This 
is one of the simplest structures admitting a complete PBG \cite{John99}.
The PC, of dimensions $8\mu m\times 8\mu m\times 8\mu m$, was placed in air and excited by a $2\mu m$-waist linearly y-polarized gaussian beam, obtained
by a total field/scattered field layer.\cite{TafloveBook} The input temporal profile, for the quasi-cw excitation
mentioned below, exhibited smooth transitions from zero to continuous wave, 
as mimicked by an {\it mnm pulse} ($m=2$, $n=10^7$). \cite{Ziolkowski01}  Its spectrum was well peaked about the carrier frequency.
The FCC lattice for a material with index $3.5$, such as Si or GaAs, has a complete band gap
around a normalized frequency $\Lambda/\lambda=0.8$, with $\lambda$ the
wavelength. To get a gap near $\lambda=1500nm$ we chose $\Lambda=1200nm$, and
the parameters of the single pole dispersion were taken as 
$\epsilon_s=11.971392$, $\omega_0=1.1\times10^{16}$ and $\gamma_0=2\times10^5$ (in MKS units), 
yielding an index $\cong3.5$ at $\lambda\cong1500nm$.
The integration domain was discretized with $dx\cong dy\cong dz\cong 30nm$ and time step $dt=0.03fs$, allowing more than $40$ points at 
each wavelength around $\lambda=1500nm$ and runs with $20000$ steps in time (the spectral resolution is of the order of $10nm$ at $\lambda=1500nm$).

\begin{figure}
\includegraphics[width=60mm]{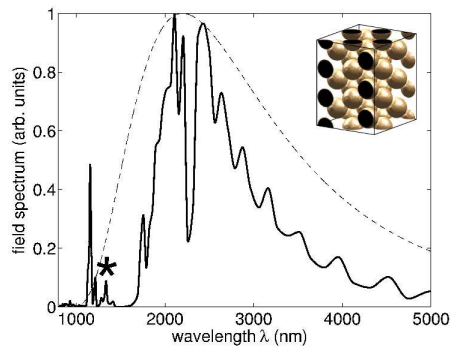}
\caption{
(Colors)Spectrum of the transmitted signal for a single-cycle low-power  excitation. The dotted line is the input spectrum
(normalized to unity).
The star marks the wavelength used to pump the OPO. A sketch of the FCC lattice is on the upper right.
\label{figuresingle}}
\end{figure}

The DOS is typically calculated by the use of plane-wave expansion,
omitting material dispersion and for infinitely extended structures (see, e.g., \cite{John99,Wang03}). Hence, it is all but straightforward the application of the standard approach to the case under consideration. For this reason,
we resorted to a time-domain approach to determine the states of the FCC-PC. A very low-power ($1nW$) single-cycle pulse \cite{Ziolkowski01} 
excited the PC (along the $\Gamma X$ direction) and the transmitted signal (its $E^y$ component) was analyzed just after it. The resulting spectrum is shown in figure \ref{figuresingle}, where the peaks correspond to concentrations of states 
(taken aside the low-frequency oscillations due to the
finitess of the structure \cite{SakodaBook}) compatibly with symmetry constraints.   
The band structure of this medium encompasses a PBG around $1500nm$ and a pseudo-gap around $2400nm$.

\begin{figure}
\includegraphics[width=60mm]{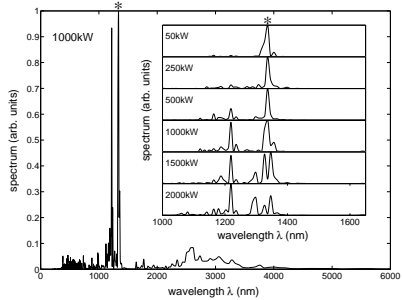}
\caption{Oscillation spectrum inside the FCC for $P_{in}=1MW$. Several excited wavelengths correspond to 
the spectral peaks in figure \ref{figuresingle}.
The asterisk marks the pump wavelength.
The insets are close-ups of the spectral interval around $1200 nm$ for different input levels.
At high powers a slight shift of the resonance peaks is due to self- and cross-phase modulation.
\label{figureSpectrum1}}
\end{figure}

In figure \ref{figureSpectrum1} we display the $E^y$ spectrum obtained in a low-symmetry point at the center of the PC, 
for a beam propagating along the direction $\Gamma X$.
Different input powers were injected with a $600fs$ quasi-cw excitation.
The pump wavelength $\lambda\cong1336nm$ corresponds to $\Lambda/\lambda=0.898$, i.e. close to a state by the (frequency) upper edge of the PBG, as indicated by the star 
in figure \ref{figuresingle}.
As visible in the insets, large output spectra are attained. 
No oscillations appear at frequencies within the PBG, with a smooth spectrum in the large wavelength region and
several peaks above the PBG upper edge. Each peak corresponds to a region dense of states,\cite{John99} and hence to an efficiently generated frequency.

\begin{figure}
\includegraphics[width=60mm]{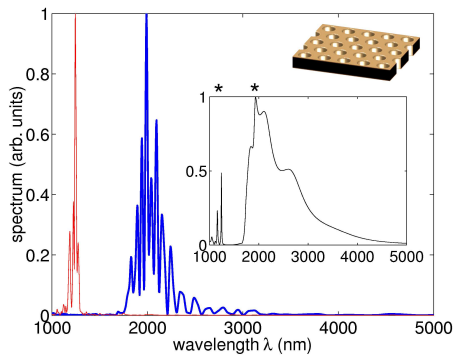}
\caption{(Colors) Oscillation spectra inside a photonic crystal slab (upper right), with generation of several wavelengths ($P_{in}=100kW$).
The thick line corresponds to a pump at $\lambda=1250 nm$, by the PBG upper edge; the thin line to $\lambda=2000 nm$, by the lower edge.
The inset shows the output spectrum for the low power ($1nW$) single cycle excitation, the stars indicate the OPO pumps. 
\label{figureslab1}}
\end{figure}

As a second representative case, we considered a photonic crystal slab with a triangular lattice ($\Lambda=450nm$, hole radius $r=135nm$),
 as experimentally investigated in \cite{Kawai01}.
The slab is $270nm$ thick, with in-plane size $2\mu m\times2\mu m$. 
It is pumped by a $0.5\mu m$-waist TE-like polarized gaussian beam incident along the $\Gamma M$ direction.
The pump wavelengths are in proximity of the upper and the lower edges of the guided-modes gap (located in the interval 1260-1750nm for the infinite non-dispersive structure), and the material constants are $\omega_0=1.1\times10^{16}$, $\gamma_0=0$, $P_0=1$, $\epsilon_s=10.9379$, and 
refractive index $n\cong3.3$ as in $Al_{0.1} Ga_{0.9} As$ at $\lambda=1800 nm$. \cite{Afromovitz74} 
The discretization was implemented with $dx\cong15nm$, $dy\cong25nm$, $dz\cong19nm$, $dt\cong0.02fs$, and 
both single cycle and cw feeding were carried out for up to $30000$ steps in time ($\cong600 fs$). 
Figure \ref{figureslab1} shows OPO spectra of the transmitted field for pump wavelengths  $\lambda=1250$
and $2000 nm$, respectively (mnm pulse, $m=5$, $n=600$), and input power $P_{in}=100kW$. 
Even at such low power, the broadening is enhanced owing to the spatial confinement afforded by a planar geometry.

In conclusion, we have analyzed optical parametric oscillations in
photonic crystal cavities. 
Through fully vectorial 3D 
numerical simulations we have demonstrated that regions of significant mode-concentration and 
high Q-factors favor the onset of parametric oscillations due to four-wave mixing gain. 
Moreover, a properly tailored DOS can favor the generation 
of specific frequencies, making nonlinear amplification possible even with isotropic (i.e. centrosymmetric) materials in microcavities.
These results pave the way to novel generations of specifically tailored parametric sources employing isotropic materials in highly integrated geometries.

We acknowledge support from INFM-``Initiative Parallel Computing'' and Fondazione Tronchetti-Provera.

\end{document}